\documentclass[12pt]{article}
\usepackage{graphicx,sectsty,amsthm,amssymb,multirow,amsmath,mathabx,comment,array}
\usepackage[round, comma, authoryear]{natbib}
\usepackage{algorithm}
\usepackage{epstopdf}
\usepackage{multirow}
\usepackage{bbm}
\usepackage{url}
\usepackage{setspace}

\sectionfont{\nohang \centering}

\makeatletter
\def \@seccntformat#1{\@ifundefined{#1@cntformat}
{\csname the#1\endcsname\quad}
{\csname #1@cntformat\endcsname}
}
\def\section@cntformat{\thesection.}
\makeatother
\theoremstyle{definition}

\usepackage{setspace}

\newcommand{\bm}[1]{\mbox{\boldmath$ #1 $\unboldmath}}

\newcommand{\figref}[1]{\mbox{figure~\ref{#1}}}

\newcommand{\secref}[1]{\mbox{section~\ref{#1}}}
\newcommand{\appref}[1]{\mbox{appendix~\ref{#1}}}

\newcommand{\theref}[1]{\mbox{theorem~\ref{#1}}}

\newcommand{\algref}[1]{\mbox{algorithm~\ref{#1}}}

\newcommand{\Figref}[1]{\mbox{Figure~\ref{#1}}}
\newcommand{\Tabref}[1]{\mbox{Table~\ref{#1}}}
\newcommand{\Secref}[1]{\mbox{Section~\ref{#1}}}
\newcommand{\Appref}[1]{\mbox{Appendix~\ref{#1}}}

\newcommand{\Algref}[1]{\mbox{Algorithm~\ref{#1}}}

\newtheorem{thm}{Theorem}

\newtheorem{obs}{Observation}
\newtheorem{defin}{Definition}

\begin{document}
\begin{center}
{\Large\bf Fast prediction of deterministic functions using sparse grid experimental designs}\\
{\Large Matthew Plumlee\footnotemark[1]}\\
{\small \footnotemark[1]Georgia Institute of Technology,  Atlanta, Georgia} \\ %{\small Georgia Institute of Technology, Atlanta, GA 30332}\\
{(\small mplumlee@gatech.edu)}\\
\end{center}

\begin{abstract}
Random field models have been widely employed to develop a predictor of an expensive function based on observations from an experiment.  The traditional framework for developing a predictor with random field models can fail due to the computational burden it requires.  This problem is often seen in cases where the input of the expensive function is high dimensional.  While many previous works have focused on developing an approximative  predictor to resolve these issues, this article investigates a different solution mechanism.  We demonstrate that when a general set of designs is employed, the resulting predictor is quick to compute and has reasonable accuracy.  The fast computation of the predictor is made possible through an algorithm proposed by this work.   This paper also demonstrates methods to quickly evaluate the likelihood of the observations and describes some fast maximum likelihood estimates for unknown parameters of the random field. The computational savings can be several orders of magnitude when the input is located in a high dimensional space.  Beyond the fast computation of the predictor, existing research has demonstrated that a subset of these designs  generate predictors that are asymptotically efficient.  This work details some empirical comparisons to the more common space-filling designs that verify the designs are competitive in terms of resulting prediction accuracy.   
\end{abstract}

Matthew Plumlee is currently a Ph.D. candidate at the H. Milton Stewart School of Industrial and Systems Engineering at the Georgia Institute of Technology.  This document is in-press at the Journal of the American Statistical Association.  A MATLAB package released along with this document is available at \url{http://www.mathworks.com/matlabcentral/fileexchange/45668-sparse-grid-designs}.

\newpage
\section{Predicting expensive functions} \label{sec:intro}
Consider a case where a deterministic output can be observed corresponding to a controllable input and the cost of an observation is expensive or at least non-negligible.    Analysis that requires a huge number of evaluations of the expensive function for different inputs can prove impractical. This paper examines a method to avoid the impracticality problem with the creation of a function that behaves similarly to the function of interest with a relatively cheap evaluation cost.  We term this cheap function a \emph{predictor} as it can closely match the output for an untried input.  The predictor can be used in place of the expensive function for subsequent analysis.

Beginning in the 1980s, research has emphasized the use of \emph{Gaussian process} models to construct predictors of the expensive function \citep{sacks1989design}.    This method, often referred to as \emph{kriging}, sprouted in geostatistics \citep{matheron1963principles} and is considered the standard approach to study expensive, deterministic functions.  A great deal of attention has been paid to this method and important variations over the last two decades because of the increased emphasis on computer simulation \citep{kennedy2001bayesian,santner2003design,higdon2008computer,gramacy2008bayesian}.  The major objectives for analysis outlined in \cite{sacks1989design} have remained basically constant: predict the output given inputs, optimize the function, and adjust inputs of the function to match observed data.  Recently, researchers have studied a fourth objective of computing the uncertainty of the output when inputs are uncertain, a topic in the broad field of uncertainty quantification. All of these objectives can be achieved through the use of a predictor, though sometimes under different names, e.g. \emph{emulator} or \emph{interpolator}.
\subsection{Predictors  and Gaussian process modeling} \label{sec:GPmodelandprediction}
As summarized in \cite{sacks1989design}, a predictor is constructed by assuming that the output, termed $y(\cdot)$, is a realization of an unknown, random function of a $d$ dimensional input $\bm x$ in a space $X \subset \mathbb{R}^d$.   One notational comment: each element in an input $\bm x$ is denoted $x^{(j)}$, i.e. $\bm x = [x^{(1)}, x^{(2)}, \ldots, x^{(d)}]$,  while sequences of inputs are denoted with subscripts, e.g. $\bm x_{1}, \bm x_{2},\ldots$.  To construct a predictor, an experiment is performed by evaluating the function for a given experimental \emph{design} (a sequence of inputs), $\mathcal{X} = \{\bm x_1, \ldots, \bm x_N\}$, creating a vector of observations $\bm y = [y(\bm x_{1}),\ldots,y(\bm x_{N})]^\mathsf{T}$.  The value of $N$ is known as the \emph{sample size} of the experimental design.  A smaller sample size represents a less expensive design.

After observing these input/output pairs, a predictor is then built by finding a representative function based on the observations.
The often adopted approach treats the unknown function as the realization of a stochastic process.   Specifically, $y(\cdot)$ is a realization of a random function $Y(\cdot)$ which has the density of a Gaussian process.  The capitalization of the output $Y(\bm x)$ indicates a random output while the lower case $y(\bm x)$ indicates the observed realization.  We denote the Gaussian process assumption on a random function $Y(\cdot)$ as
\[Y(\cdot) \sim GP(\mu (\cdot),C(\cdot,\cdot)),\]
where $\mu(\cdot)$ is the mean function and $C(\cdot,\cdot)$ is a function such that $C(\bm x_1, \bm x_2) = \operatorname{cov}(Y(\bm x_1), Y(\bm x_2))$ for all possible $\bm x_1, \bm x_2 \in X$.

A typical assumption on the covariance structure is a separable covariance, defined as $C(\bm x_1, \bm x_2) = \prod_{i=1}^d C_i(x_1^{(i)},x_2^{(i)})$ for all $\bm x_1, \bm x_2 \in X$.  The functions $C_i$ are covariance functions defined when the input is one dimensional. The value of $C_i(x,x')$ is proportional to the covariance between two outputs corresponding to inputs where only the $i$th input differs from $x$ to $x'$.   The results in this paper require this covariance structure to hold, but no other assumptions are needed for $\mu(\cdot)$ and $C(\cdot,\cdot)$.  \Secref{sec:gen_prediction} discusses estimating the aforementioned mean and covariance functions using the observations when they are unknown.  For now, we consider these functions known for simplicity of exposition.

Our goal is to predict an unobserved output at an untried input $\bm x_0$ given $\bm Y := [Y(\bm x_{1}),\ldots,Y(\bm x_{N})]^\mathsf{T} = \bm y$.  The commonly used predictor of $y (\bm x_0)$ is
\begin{equation}
\hat{y} \left(\bm x_0\right) = \mu(\bm x_0)+ \bm \sigma^\mathsf{T}(\bm x_0) \bm w, \label{eq:GP_pred}
 \end{equation} where $\bm w \in \mathbb{R}^N$ is a vector of weights and $\bm \sigma^\mathsf{T}(\bm x_0) = [C(\bm x_0,\bm x_1),\ldots,C(\bm x_0,\bm x_N)]$.    In general, $\bm w$ is given by the following relation
\[\bm w = \bm \Sigma^{-1} \left(\bm y - \bm \mu\right),\]
where $\bm \mu = [\mu(\bm x_{1}),\ldots,\mu(\bm x_{N})]^\mathsf{T}$ and $\bm \Sigma$ is the $N \times N$ covariance matrix where the element in the $i$th row and $j$th column is $C(\bm x_i,\bm x_j)$.  This predictor, $\hat{y} \left(\bm x_0\right)$, is commonly used because it is both the mean and median of the predictive distribution of $Y (\bm x_0)$ given $\bm Y = \bm y$. This property implies $\hat{y} \left(\bm x_0\right)$ is optimal among the class of both linear and nonlinear predictors of $y(\bm x_0)$ with respect to the quadratic and absolute loss functions.
\subsection{Focus of the paper}\label{sec:focus}
The above approach, when applied in a direct manner, can become intractable because the inversion of the covariance matrix $\bm \Sigma$ is an expensive operation in terms of both memory and processing.   Direct inversion can also induce numerical errors due to limitations of floating point mathematical computations \citep{wendland2005scattered,haaland2011accurate}.   Previous research has focused on changing the matrix $\bm \Sigma$ to a matrix that is easier to invert, therefore making the computation of $\bm w$ faster \citep{furrer2006covariance,cressie2008fixed,banerjee2008gaussian}.  We term this an \emph{approximation} because this can degrade predictive performance, though sometimes only slightly.

In this work, we forgo approximations and investigate a new approach to resolve this problem: by restricting ourselves a general class of designs, accurate  non-approximative predictors can be with found with significantly less computational expense.  This class of experimental designs is termed \emph{sparse grid designs} and is based on the structure of eponymic interpolation and quadrature rules.  Sparse grid designs \citep{smolyak1963quadrature} have been used with in conjunction with polynomial rules \citep{wasilkowski1995explicit,barthelmann2000high, xiu2006high,nobile2008sparse,xiu2010numerical}, but these designs have not gained popularity among users of random field models.  Here, we encourage the use of sparse grid designs by demonstrating computational procedures to be used with these designs where the predictor can be computed very quickly.

\Secref{sec:back_designs} will briefly describe two broad types of existing designs and identify deficiencies of those existing types.  Section \ref{sec:designs} will explain the definition of sparse grid designs and then the following sections will discuss three important topics:
 \begin{itemize}
 \item Section \ref{sec:sg_infr} explains how we can exploit the structures used in building sparse grid designs to achieve extreme computational gains when building the predictor.
     Our algorithm computes $\bm w$ by inverting several small matrices versus one large matrix.  This algorithm is derived from the result that $\hat{y} \left(\bm x_0\right)$ can be written as the tensor product of linear operators, see \theref{thm:optimal_sg} in \appref{app:BLUP_Lin}.
\item \Secref{sec:gen_prediction} goes on to demonstrate that we can estimate unknown parameters of the random field with similar computational quickness.  Of note is \theref{thm:det}, which gives an expression for the determinant of the matrix $\bm \Sigma$ that can be evaluated quickly.
 \item  \Secref{sec:simu_comp} illustrates that sparse grid designs perform well even when the input is high dimensional.  We conduct empirical comparisons that demonstrate good performance of these designs which supports the positive asymptotic arguments proven previously \citep{temlyakov1987approximate}.
     \end{itemize}
Section \ref{sec:conc} will offer some discussions on the role of these designs and the creation of optimal sparse grid designs.
\section{Space-filling and lattice designs} \label{sec:back_designs}
\begin{figure}[tbp]
\begin{center}
\includegraphics[width=6.5 in]{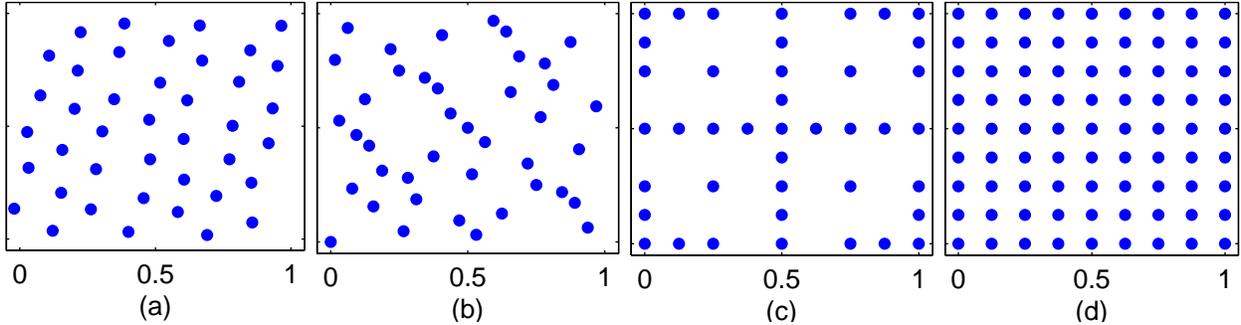}
\caption{Examples of 2-dimensional designs: (a) A $41$ point %maximin
Latin-hypercube design. (b) The first $41$ points in the Sobol sequence. (c) A $41$ point sparse grid design. (d) An $81$ point lattice design.  Details of the construction of the sparse grid design in (c) are given in appendix \ref{app:componet}.}
\label{fig:grid_compare}
\end{center}
\end{figure}

This section will briefly discuss existing research on \emph{space-filling} and \emph{lattice} designs.  The space-filling category includes the popular Latin hypercube designs.  Lattice designs are a specific class of designs where each design is a \emph{Cartesian product} of one dimensional designs.  Visual examples are given in \figref{fig:grid_compare} and they are contrasted with an example of a sparse grid design which will be explained in \secref{sec:designs}.
\subsection{Space-filling designs: Efficient predictors but difficult computation}
Current research has emphasized the design of points that are \emph{space-filling} (see \figref{fig:grid_compare} (a) and (b)).
Designs of this type are often scattered, meaning they are not necessarily located on a lattice.
The major focus has been on Latin hypercube designs \citep{mckay1979comparison}, seen in \figref{fig:grid_compare} (a), and research has produced a swell of variations, e.g. \cite{tang1993orthogonal,owen1994controlling,morris1995exploratory,ye1998orthogonal,joseph2008orthogonal}.  These designs have been shown to perform well in many prediction scenarios and are often considered the standard method of designing computer experiments for deterministic functions.

However, space-filling designs experience significant difficulties when the input is high dimensional, i.e. $d > 3$.  In these cases,  one  requires a large sample size $N$ to develop an accurate predictor.  This in turn makes the matrix $\bm \Sigma$ very large, meaning $\bm w$ is difficult to compute through inversion of $\bm \Sigma$.  This has motivated the research into approximate predictors discussed in \secref{sec:focus} that can be used with space-filling designs.
\subsection{Lattice designs: Easy computation but inefficient predictors} \label{subsec:lattice_des}
One of the simplest forms of an experimental design is a \emph{lattice} design, also known as a grid.  This is defined as $\mathcal{X} = \mathcal{X}_1 \times \mathcal{X}_2 \times \ldots \times \mathcal{X}_d$ where each $\mathcal{X}_i$ is a set of one dimensional points we term a \emph{component} design. For a set $A$ and $B$, the Cartesian product, denoted $A \times B$, is defined as the set of all ordered pairs $(a,b)$ where $a \in A$ and $b \in B$.  If the number of elements in $\mathcal{X}_i$ is $n_i$, then the sample size of a lattice design is  $\prod_{i=1}^d n_i$.

Let the covariance be as stated in \secref{sec:GPmodelandprediction},  $C(\bm x_1, \bm x_2) = \prod_{i=1}^d C_i(x_1^{(i)},x_2^{(i)})$.  When a lattice design is used, the covariance matrix takes the form of a \emph{Kronecker product} of matrices: $\otimes_{i=1}^d \bm S_i$, where $\bm S_i$ is a matrix composed of elements $C_i(x,x')$ for all $x,x' \in\mathcal{X}_i$.  A useful property of Kronecker products can be derived using only the definition of matrix multiplication and the commutativity of scalar multiplication: if $\bm A= \bm C \otimes \bm E$ and $\bm B = \bm D \otimes \bm F$ then  $\bm A\bm B = \bm C \bm D \otimes \bm E\bm F$ (when matrices are appropriately sized).  This immediately implies that if $\bm C$ and $\bm E$ are both invertible matrices, $\bm A^{-1} = \bm C^{-1} \otimes \bm E^{-1}$.  Thus, if a lattice design is used, \[\bm w = \left(\otimes_{i=1}^d \bm S_i^{-1} \right) \left(\bm y - \bm \mu \right),\]
which is an extremely fast algorithm because $\bm S_i$ are $n_i$ sized matrices.  Many authors have noted the power of using lattice designs for fast inference for these types of models \citep{o1991bayes,bernardo1992integrated}. Say that we have a symmetric design where $\mathcal{X}_i = \mathcal{X}_j$ for all $i$ and $j$.   Computing $\bm w$  requires inversion of $N^{1/d} \times N^{1/d}$ sized matrices which are much smaller than the $N \times N$ sized matrix $\bm \Sigma$.  Because inversion of an $N \times N$ size matrix requires $\mathcal{O}(N^3)$ arithmetic operations, inverting multiple small matrices versus one large one yields significant computational savings.

While lattice designs are extremely simple and result in fast-to-compute predictors, these are wholly impractical for use in high dimensions.  First, lattices are grossly inefficient as experimental designs when the dimension is somewhat large ($d>3$), which will be demonstrated in \secref{sec:simu_comp}.   Also, the sample size of a lattices designs, $\prod_{i=1}^d n_i$, is extremely inflexible regardless of the choice of $n_i$.  At minimum $n_i=2$, and then even for a reasonable number of dimensions the size of the design can become quite large. When $d= 15$ the smallest possible design size is over $30,000$.
\section{Sparse grid designs} \label{sec:designs}
This section will discuss the construction of sparse grid experimental designs which are closely associated with sparse grid interpolation and quadrature rules.    To build these designs first specify a nested sequence of one dimensional experimental designs for each $i=1,\ldots,d$ denoted $\mathcal{X}_{i,j}$, where  $\mathcal{X}_{i,j} \subseteq \mathcal{X}_{i,j+1}, j = 0,1,2, \ldots$, and $\mathcal{X}_{i,0} = \emptyset$.  Designs defined for a single dimension, e.g. $\mathcal{X}_{i,j}$, are termed \emph{component designs} in this work.  The nested feature of these sequences is important for our case.  The general literature related to sparse grid rules does not require this property.  Here, it is necessary for the stated results to hold.

\begin{figure}[tbp]
\begin{center}
\includegraphics[width=5 in]{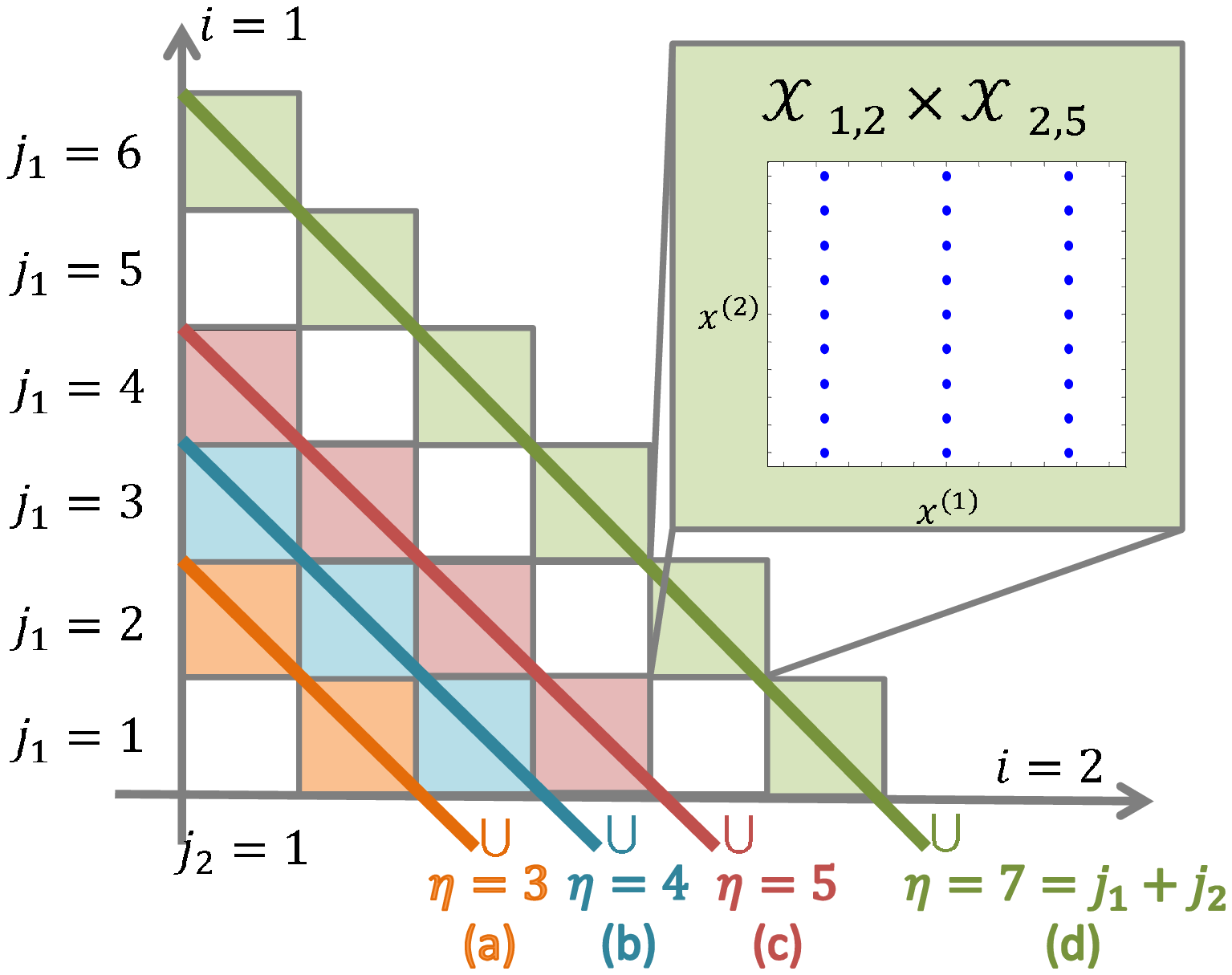}
\caption{Diagram of the construction of the two dimensional designs seen in \figref{fig:grid_exmp}. Each box represents $\mathcal{X}_{1,j_1} \times  \mathcal{X}_{2,j_2}$. The dark lines pass through lattice designs creating the union of the sets featured in \figref{fig:grid_exmp}.}
\label{fig:const_illistr}
\end{center}
\begin{center}
\includegraphics[width=6.5 in]{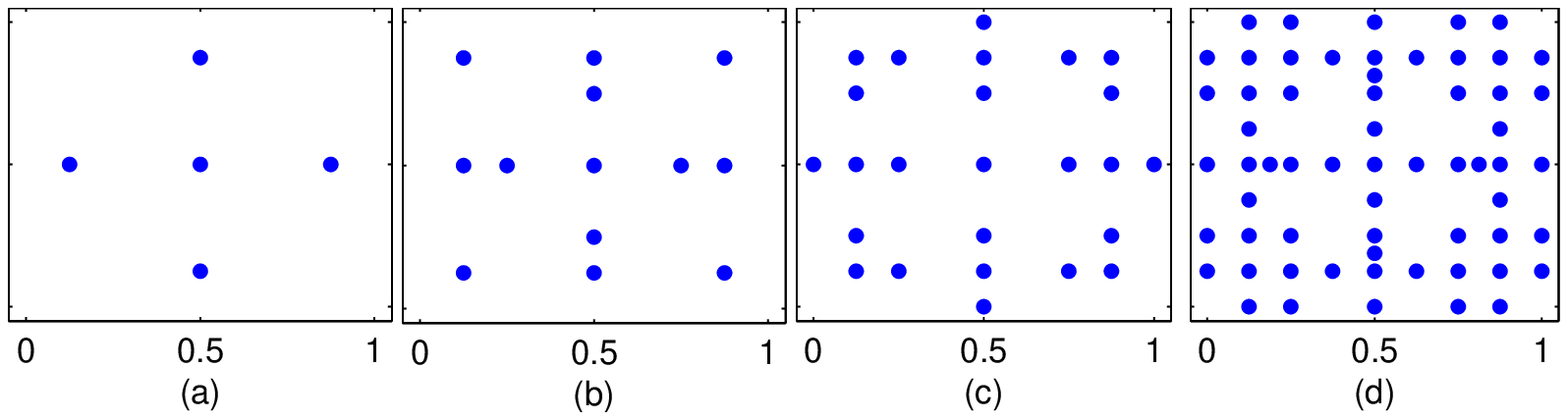}
\caption{Sparse grid designs associated with \figref{fig:const_illistr} where $d=2$ and $\eta$ = 3 (a), 4 (b), 5 (c), and 7 (d). The details of the component designs used for this figure can be seen in appendix \ref{app:componet}.}
\label{fig:grid_exmp}
\end{center}
\end{figure}

Sparse grid designs are therefore defined as
\begin{equation}
\mathcal{X}_{SG} (\eta) = \bigcup_{{\scriptstyle { \vec{j}} \in \mathbb{G}(\eta)}} \mathcal{X}_{1,j_1} \times \mathcal{X}_{2,j_2}\times \cdots \times \mathcal{X}_{d,j_d}, \label{eq:num_points}
\end{equation}
where $\eta \geq d$ is an integer that represents the level of the construction and $\mathbb{G}(\eta) = \left\{\vec{j} \in \mathbb{N}^d | \sum_{i=1}^d j_i = \eta\right\}$.  Here we use the overhead arrow to distinguish the vector of indices, $\vec{j} = [j_1, \ldots, j_d]$ from a scalar index.  Increasing the value of $\eta$ results in denser designs.   \Figref{fig:const_illistr} illustrates the construction of the two dimensional designs seen in \figref{fig:grid_exmp}. The details of the component designs can be seen in appendix \ref{app:componet}.

Unlike many other design alternatives, sparse grid designs are not defined via a given sample size.  The sample size of the resulting sparse grid design is a complicated endeavor to compute \emph{a-priori}.
After the dimension $d$ and level of construction $\eta$, a major contributing factor to the sample size,  $N_{SG}(\eta) := \#\mathcal{X}_{SG} (\eta)$, is the sizes of the component designs.  The sample size of a sparse grid design is given by
\[N_{SG} (\eta) = \sum_{\scriptstyle {\vec{j}} \in \mathbb{J}(\eta)} \prod_{i=1}^d \#\mathcal{X}_{i,j}- \#\mathcal{X}_{i,j-1},\]
  where $\mathbb{J}(\eta) = \left\{\vec{j} \in \mathbb{N}^d | \sum_{i=1}^d j_i \leq \eta\right\}$.  \Tabref{tab1} presents some shortcut calculations of the sample size along with some bounds when $\#\mathcal{X}_{i,j} = \#\mathcal{X}_{k,j} = h(j)$ for all $i$ and $k$.
%\singlespacing
\begin{table}
\centering
\caption{Sample size of sparse grid designs with level of construction $\eta$, dimension $d$ and $\#\mathcal{X}_{i,j} = h(j)$ for all $i$. The values of $c$ and $c_0$ are some constant integers bigger than zero.  The last line is from \cite{wasilkowski1995explicit}.}\label{tab1}
\begin{tabular}{|l| l| l| } \hline
$h(j), j > 0$ &  $N_{SG}(\eta)$  & Bound on $N_{SG}(\eta)$ \\  \hline
 \vspace{2pt} $c j$ &  $c^d {\eta \choose d}$  & $(c \eta)^d/d! $  \vspace{2pt} \\
 \vspace{2pt}$c(j - 1) + 1$ &  $\sum_{k=0}^{\min(d,\eta-d)} c^k {d\choose k} {\eta-d \choose k}$  & $c^{\eta-d}  {\eta \choose d} $ if  $\eta \leq 2 d$ \vspace{2pt}\\
 \vspace{2pt} $c_0( c^j - 1)$ &  $ c_0^d (c -1)^d \sum_{j=0}^{\eta-d} c^j {j + d -1 \choose d-1} $  & $c_0^d \left(c - 1\right)^{d-1} c^{\eta-d+1} {\eta - 1 \choose d-1}$  \vspace{2pt} \\ \hline
\end{tabular}
\end{table}

The proper selection of the points in the component designs is essential to achieving good performance of the overall sparse grid design.  Establishing good component designs  can lead to a good sparse grid design, but interaction between dimensions is an important consideration.
\section{Fast prediction with sparse grid designs} \label{sec:sg_infr}
This section will propose an algorithm that shows the major advantage of sparse grid designs: the availability of fast predictors.  \Appref{app:BLUP_Lin} justifies the proposed algorithm by describing a predictor in the form of a tensor product of linear maps and then \theref{thm:optimal_sg} demonstrates that conjectured predictor is the same as $\hat{y} (\bm x_0)$.

Here we show how to build the weight vector, $\bm w$, by inverting covariance matrices associated with the component designs which are relativity small compared to $\bm \Sigma$.
Therefore, our proposed method results in a faster computation of $\bm w$ than a method that computes $\bm w$ from direct inversion of $\bm \Sigma$ when $N$ is large.    This is the same mechanism that is used to construct fast predictors with lattice designs.  But unlike lattice designs, we show sparse grid designs perform well in cases where the input is high dimensional in  \secref{sec:simu_comp}.

\Algref{alg:w_comp} lists the proposed algorithm for computing $\bm w$.  In the algorithm,  the matrices  $\bm S_{i,j}$  are composed of elements $C_i(x, x')$, for all $x, x' \in \mathcal{X}_{i,j}$.   Also,  the vectors $\bm y_{{ \vec{ j}}}$, $\bm \mu_{{ \vec{ j}}}$ and $\bm w_{{ \vec{ j}}}$ denote subvectors of $\bm y$, $\bm \mu$ and $\bm w$ at  indices corresponding to  $\mathcal{X}_{1,j_1} \times \mathcal{X}_{2,j_2} \times \ldots \times \mathcal{X}_{d,j_d}$ for all $\vec{j} \in \mathbb{J}(\eta)$.    \setcounter{algorithm}{0}
\begin{algorithm}[h]
\centering
\begin{minipage}{6in}
\caption{Proposed algorithm for the fast computation of $\bm w$ when the design is $\mathcal{X}_{SG}(\eta)
$.  Here, $a(\vec{j}) = \left(-1\right)^{\eta - {\scriptscriptstyle|\vec{j}|}} {d-1 \choose \eta-{\scriptscriptstyle|\vec{j}|}}$ and  $\mathbb{P}(\eta)  = \left\{ \vec{j} \in \mathbb{N}^d | \max(d,\eta-d+1) \leq \sum_{i=1}^d j_i \leq \eta \right\}$.} \label{alg:w_comp}
\begin{tabbing}
   \enspace Initialize $\bm w = \bm 0$ \\
   \enspace For all $ \vec{j} \in \mathbb{P}(\eta)$   \\
    \qquad  $ \bm w_{\vec{j}} = \bm w_{\vec{j}}+ a(\vec{j}) \left(\bigotimes_{i=1}^d \bm S_{i,j_i}^{-1}\right)  \left(\bm y_{\vec{j}}-\bm \mu_{{ \vec{ j}}}\right)$
\end{tabbing}
\end{minipage}
\end{algorithm}

Another important feature of predictors in general is the presence of a predictive variance,   \[\mathbb{E}_{\bm Y=\bm y} \left(\hat{y}(\bm x_0)-Y(\bm x_0)\right)^2 =C(\bm x_0, \bm x_0) - \bm \sigma^\mathsf{T}(\bm x_0) \bm \Sigma^{-1} \bm \sigma(\bm x_0),\]
 where the subscript $\bm Y=\bm y$ on the expectation implies we condition on that case. As noted before, computation of $\bm \Sigma^{-1}$ is an undesirable operation.  Luckily, this operation can be avoided by using sparse grid designs.  As demonstrated in \appref{app:SG_MSPE}, when employing a sparse grid design the predictive variance is given by
\begin{equation}
\mathbb{E}_{\bm Y=\bm y} \left(\hat{y}(\bm x_0)-Y(\bm x_0)\right)^2 = C(\bm x_0,\bm x_0) - \sum_{{ \vec{j}} \in \mathbb{J}(\eta)} \prod_{i=1}^d \Delta_{i,j_i}(\bm x_0), \label{eq:sg_error}
\end{equation}
where $\Delta_{i,j}(\bm x_0) = \varepsilon_{i,j-1}(\bm x_0) - \varepsilon_{i,j}(\bm x_0)$ and $\varepsilon_{i,j}$ is defined as the expected squared prediction error in one dimension with covariance $C_i$ and design $\mathcal{X}_{i,j}$.  After substituting known relations, we have that
\[\varepsilon_{i,j}(\bm x_0)  = C_i(x_0^{(i)},x_0^{(i)})  - \bm s_{i,j}^\mathsf{T}( x_0^{(i)}) \bm S_{i,j}^{-1}  \bm s_{i,j}(x_0^{(i)}),\]
where the elements of the vector $\bm s_{i,j} (x_0^{(i)})$ are $C_i(x_0^{(i)},x)$  for all $x \in\mathcal{X}_{i,j}$.
\obs{Some exact comparisons are helpful to understand how fast  \algref{alg:w_comp}'s computation of $\bm w$ is compared to the traditional method.  Using the same settings as \secref{sec:mse_comp} with $\eta = 14$ and $d=10$, the computation of $\bm w$ took $.35$ seconds with the proposed algorithm and the traditional method of computing $\bm w$ by inverting $\bm \Sigma$ took $40$ seconds.  In a much larger example with  $N =467,321$ ($\eta = 73$ and $d=70$), computing $\bm w$ using the proposed algorithm took $14.7$ seconds.   If we assume the cost of prediction scales at the rate of matrix inversion, $N^3$, then computing $\bm w$ using the traditional method of inverting $\bm \Sigma$ with a design size of $467,321$ would take approximately 81 days to compute.}
\section{Fast prediction with unknown parameters}\label{sec:gen_prediction}
The previous section assumed that both mean, $\mu(\cdot)$, and covariance, $C(\cdot,\cdot)$, are exactly known.  This is often not assumed in practical situations.  Instead, these functions are given general structures with unknown parameters which we denote $\theta$.   Two major paradigms exist for prediction when $\theta$ is unknown: (i) simply use an estimate for $\theta$ based on the observations and predict using (\ref{eq:GP_pred}) or (ii)  Bayesian approaches \citep{santner2003design}.   For either method, the typical formulae require computation of both the determinant and inverse of $\bm \Sigma$, which are costly when $N$ is large.  This section develops the methods to avoid these computations.  For expositional simplicity, this section will outline the first method and leave the full Bayesian method for future work.

The estimate of $\theta$ we consider will be the \emph{maximum likelihood estimate} (MLE), which is denoted $\hat{\theta}$.  We therefore term the predictor that uses this estimate as the \emph{MLE-predictor}, which will be used for comparisons in \secref{sec:deter_fun}.  We first explain the typical general structures of $\mu(\cdot)$ and  $C(\cdot,\cdot)$ in \secref{subsec:gen_setting} and then we describe the traditional forms of the estimate  $\hat{\theta}$ and problems with them in \secref{subsec:trad_MLEs}.  \Secref{subsec:MLEs} then explains fast methods to find $\hat{\theta}$ in this setting.
\subsection{General setting} \label{subsec:gen_setting}
The structures of $\mu$, $C$ and $\theta$ in this section are borrowed from \cite{santner2003design} and are widely employed. We assume that the mean is a linear combination of $p \geq 1$ basis functions, $f_1(\cdot), \ldots, f_p (\cdot)$, and the covariance function is scaled such that $C(\cdot, \cdot) = \sigma^2 R(\cdot, \cdot ;\phi)$, where $R(\bm x_1, \bm x_2;\phi) =  \prod_{i=1}^d  R_i (x_1^{(i)}, x_2^{(i)} ;\phi)$ is a correlation function and $\sigma^2$ represents the variance of $y (\bm x) - \mu (\bm x)$.  The parameter $\phi$ is a general parameter or group of parameters that can represent unknown aspects of $R(\cdot, \cdot ;\phi)$ that affect the lengthscale and differentiability of the realized response $y(\cdot)$.  We now have the following case
\[Y(\cdot) \sim GP \left( \sum_{k =1}^p \beta_k f_k (\cdot), \sigma^2 R(\cdot,\cdot ; \phi) \right),\]
where $\theta = \{\beta_1,\ldots,\beta_p, \sigma^2, \phi\}$ is the set of unknown parameters.
\subsection{Traditional computation of the MLE} \label{subsec:trad_MLEs}
The logarithm of the probability density of the observations $\bm y$ with $\theta = \{\beta_1,\ldots,\beta_p, \sigma^2, \phi\}$, called the \emph{log-likelihood}, is given by (up to a constant)
\[L(\bm \beta, \sigma^2, \phi) = -\frac{1}{2}\left(N\log(\sigma^2) + \log |\bm R_\phi| + \left(\bm y -\bm F \bm \beta \right)^{\mathsf{T}} \bm R_\phi^{-1} \left(\bm y -\bm F \bm \beta \right)/\sigma^2\right),\]
where $|\bm A|$ represents the determinant of a matrix $\bm A$, $\bm R_\phi$ is the $N \times N$ correlation matrix of $\bm y$ when parameter $\phi$ is used,  $\bm \beta =  [\beta_1,\ldots,\beta_p]^\mathsf{T}$, and  \[\bm F = \left[ \begin{array}{ccc}
f_1(\bm x_1)& \ldots& f_p (\bm x_1) \\
f_1(\bm x_2)& \ldots& f_p (\bm x_2) \\
\vdots & \vdots & \vdots\\
f_1(\bm x_N)& \ldots& f_p (\bm x_N)\end{array}\right]. \]
Our goal is to solve the optimization problem
\[\hat{\theta} = \operatorname{argmax}_{\beta,\sigma^2,\phi} L(\bm \beta, \sigma^2, \phi). \]

There are closed form maximum likelihood estimates for both $\bm \beta$ and $\sigma^2$ given $\phi$ which we denote $\hat{\bm \beta}_{\phi}$ and $\hat{\sigma}_{\phi}^2$.  They are
\[\hat{\bm \beta}_{\phi} = \left(\bm F^\mathsf{T} \bm R_\phi^{-1} \bm F\right)^{-1}  \bm F^\mathsf{T} \bm R_\phi^{-1} \bm y\]
and
\[\hat{\sigma}_{\phi}^2 = N^{-1}\left(\bm y-\bm F \hat{\bm \beta}_{\phi} \right)^\mathsf{T}\bm R_{\phi} ^{-1}\left(\bm y-\bm F \hat{\bm \beta}_{\phi} \right).\]   Then, $\phi$ is found by generic numerical maximization, i.e.
\[\hat{\phi} = \operatorname{argmax}_{\phi} L(\hat{\bm \beta}_\phi, \hat{\sigma}^2_\phi, \phi).\]

The problem with using these methods directly is that  $\hat{\bm \beta}_{\phi}$ and $\hat{\sigma}_{\phi}^2$ require inversion of the $N \times N$ matrix $\bm R_\phi$.  Additionally, $L(\hat{\bm \beta}_\phi, \hat{\sigma}_\phi^2, \phi)$ still contains the term $\log |\bm R_\phi|$, which is often as cumbersome as finding $\bm R_\phi^{-1}$.  The remainder section proposes alternatives to these methods that are faster.   Specifically, we will be able to compute $\hat{\bm \beta}_{\phi}$, $\hat{\sigma}_{\phi}^2$ and $\log |\bm R_\phi|$ without ever storing or operating directly on $\bm R_\phi$.
\subsection{Proposed fast computation of the MLE} \label{subsec:MLEs}
To introduce our fast-to-compute maximum likelihood estimate, we first describe a generalization of \algref{alg:w_comp} seen in \algref{alg:F_tilde}. \Algref{alg:F_tilde} computes \[Q\left(\bm A;   \otimes_{i=1}^d C_i\right) := \bm \Sigma^{-1} \bm A,\] where $\bm A$ is any $N \times m$ matrix and $m$ is any positive integer.   The notation ``$\otimes_{i=1}^d C_i$'' implies we have a separable covariance with each covariance function being represented by $C_i(\cdot,\cdot)$.  The computations in \algref{alg:F_tilde} do not require the direct inversion of $\bm \Sigma$ and therefore avoid the major computational problems of the traditional method.  The validity of \algref{alg:F_tilde} is implied by the validity of \algref{alg:w_comp}.%\prod\nolimits_{i=1}^d
\begin{algorithm}[h]
\centering
\begin{minipage}{6in}
\caption{Fast computation of $Q(\bm A; \otimes_{i=1}^d C_i) = \bm \Sigma^{-1} \bm A$ when the design is $\mathcal{X}_{SG}(\eta)$ and $\bm A$ is any $N \times m$ matrix where $m$ is any positive integer.  The notation ``$\otimes_{i=1}^d C_i$'' implies we have a separable covariance with each covariance function being represented by $C_i(\cdot,\cdot)$.  The notation $\bm A_{\vec{j}, \cdot}$ means the matrix with rows that correspond to $\mathcal{X}_{1,j_1} \times \mathcal{X}_{2,j_2} \times \ldots \times \mathcal{X}_{d,j_d}$ and all columns of $\bm A$. \Secref{sec:sg_infr} defines
$\mathbb{P}(\eta)$, $a(\vec{j})$, and $\bm S_{i,j}$.} \label{alg:F_tilde}
\begin{tabbing}
   \enspace Initialize $\tilde{\bm A}$ as an $N \times m$ matrix with all $0$ entries.  \\
   \enspace For all $ \vec{j} \in \mathbb{P}(\eta)$   \\
    \qquad  $\tilde{\bm A}_{\vec{j},\cdot} = \tilde{\bm A}_{\vec{j},\cdot}+ a(\vec{j}) \left(\bigotimes_{i=1}^d \bm S_{i,j_i}^{-1}\right)  \bm A_{\vec{j}, \cdot}$\\
\enspace Output $\tilde{\bm A}$.
\end{tabbing}
\end{minipage}
\end{algorithm}

Now we can establish our maximum likelihood estimates for $\bm \beta$ and $\sigma^2$ that do not require inversion of the $N \times N$ matrix $\bm R_\phi$.  For a given $\phi$,
\[\hat{\bm \beta}_\phi = \left(\left[Q(\bm F; \otimes_{i=1}^d R_i(\phi))\right]^\mathsf{T} \bm F\right)^{-1}  \left[Q(\bm F;\otimes_{i=1}^d R_i(\phi))\right]^\mathsf{T}  \bm y\]
and
\[\hat{\sigma}_\phi^2 = N^{-1} \left[Q(\bm y-\bm F \hat{\bm \beta}_\phi; \otimes_{i=1}^d R_i(\phi))\right]^\mathsf{T} \left(\bm y-\bm F \hat{\bm \beta}_\phi\right).\]

The last step to find the MLE requires maximization of $L(\hat{\bm \beta}_\phi, \hat{\sigma}_\phi^2, \phi)$ with respect to $\phi$.  This expression contains the term $\log |\bm R_\phi|$ which,  as mentioned before, is expensive to compute.  Therefore, we demonstrate the following theorem related to the expression of the determinant that only involves determinants of component covariance matrices, $\bm  S_{i,j}$.  The proof lies in the appendix.
\begin{thm}\label{thm:det}
If $\mathcal{X} = \mathcal{X}_{SG}(\eta)$, then
\[\log |\bm \Sigma| = \sum\nolimits_{{ \vec{j}} \in \mathbb{J}(\eta)} \sum\nolimits_{i=1}^d \left( \log \left|\bm S_{i,j_i}\right| - \log \left|\bm  S_{i,j_i -1}\right|\right)\cdot  \prod\nolimits_{k \neq i} \#\mathcal{X}_{k,j_k}-\#\mathcal{X}_{k,j_k-1}\]
where $\left|\bm S_{i,0}\right| := 1$ for all $i$.
\end{thm}
By using $R(\bm x_1, \bm x_2;\phi) =  \prod_{i=1}^d  R_i (x_1^{(i)}, x_2^{(i)} ;\phi)$ as the covariance function in the formula in the above theorem, we gain an expression for $\log |\bm R_\phi|$ without directly computing the determinant of an $N \times N$ matrix.  Once $\hat{\phi}$ is found, this gives us $\hat{\theta} = \{\hat{\bm \beta}_{\hat{\phi}},\hat{\sigma}_{\hat{\phi}}^2,\hat{\phi}\}$.
\section{Prediction performance comparisons} \label{sec:simu_comp}
Thus far, this paper has established that we can build predictors quickly when sparse grid designs are used.  However, an issue of critical importance is how well the resulting predictors perform.   This section seeks to compare the predictive performance resulting from sparse grid designs to the more common designs discussed in \secref{sec:back_designs}.  Our core findings can be summarized as follows: (i) both sparse grid and space-filling designs outperform lattice designs, (ii) sparse grid designs appear competitive with space-filling designs for smooth functions and inferior to space-filling designs  for very rough functions,  and (iii) the time taken to find the MLE-predictor using sparse grid designs can be orders of magnitude less than the time taken using the traditional methods.

Before we begin numerical comparisons, it might be helpful to take a historical look at sparse grid designs.  The prevalence of sparse grid designs in the numerical approximation literature can be owed to the demonstrated efficiency of the designs even when the input is of high dimension.  It has been shown if $X = [0,1]^d$,  using sparse grid designs with component designs of the form $\mathcal{X}_{i,j} = \{1/2^{j},\ldots,(2^{j}-1)/2^{j}\}$ is an asymptotically efficient design strategy under the symmetric separable covariance structure \citep{temlyakov1987approximate,wozniakowski1992average,ritter1995multivariate}.  These designs are also known as \emph{hyperbolic cross points}. The key point discovered in the previous analysis is that sparse grid designs are asymptotically efficient regardless of dimension and lattice designs become increasingly inefficient as the dimension grows large.  Therefore, we anticipate that sparse grid designs outperform lattices in high dimensions.

The sparse grid designs used in this section were constructed from component designs that are symmetric across dimensions and details of the component designs are in appendix \ref{app:componet}.  These appeared to be at least competitive if not superior to hyperbolic cross points in a simulation study comparable to \secref{sec:mse_comp}.  The space-filling designs were constructed by using the scrambled Sobol sequence described in \cite{matouvsek1998sub}. Maximin Latin hypercube designs that were generated via the R package \verb lhs  produced inferior distance metrics for large sample sizes but the same conclusions as the ones presented in this section.  The lattice design designs used for comparison in \secref{sec:mse_comp} were $\{1/4,3/4\}^{10}$, $\{0,1/2,1\}^{10}$, and $\{0,1/3,2/3,1\}^{10}$.
\subsection{Comparison via average prediction error} \label{sec:mse_comp}
\begin{figure}[tbp]
\begin{center}
\includegraphics[width=6.5 in]{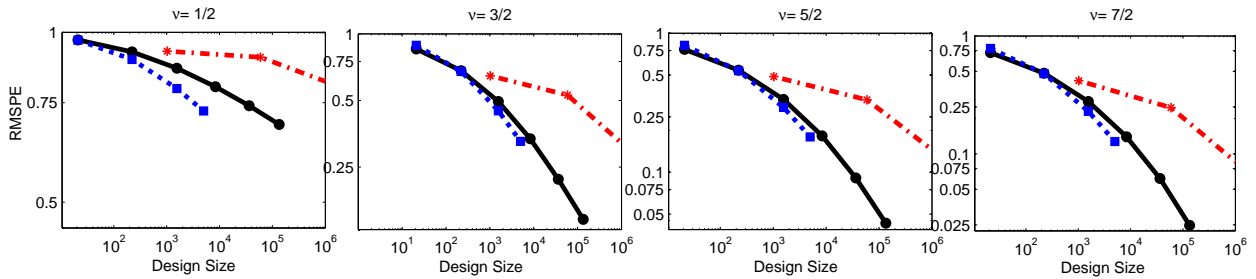}
\caption{Root mean square prediction errors (RMSPE) associated with  sparse grid designs (solid), space-filling designs (small dashes), and lattice designs (dashed-dotted) for the simulation discussed in \secref{sec:mse_comp}.  The random fields are located in $[0,1]^{10}$ and defined with a Mat\'{e}rn covariance function where $\phi = .75$ and  $\nu$ varies. }
\label{fig:smooth_compare}
\end{center}
\end{figure}

This subsection will investigate the mean square prediction error resulting from various experimental designs when the mean and covariance structures are known.  This can be thought of as the average mean squared prediction error over all possible sample paths, $y(\cdot)$, drawn from a Gaussian process with a specified covariance function. Furthermore, we seek to examine the impact of the smoothness of $y (\cdot)$ on the effectiveness of the design strategies.  To allow for the introduction of varying levels of smoothness, this section will use the Mat\'{e}rn class of covariance functions,
\begin{equation}
C_i(x,x') = \frac{1}{2^{\nu-1} \Gamma\left(\nu \right)} \left(
\sqrt{2 \nu} h  \right)^\nu \mathcal{K}_{\nu} \left(
\sqrt{2 \nu}h \right), \label{eq:matern}
\end{equation} where $\mathcal{K}_{\nu}$ is the modified Bessel function of order $\nu>0$ and $h = |x - x'|/ \phi$.  The use of this covariance class allows us to independently adjust a smoothness parameter $\nu$, where the sample paths are $\lceil\nu-1\rceil$ times differentiable \citep{handcock1993bayesian}.  For simplicity, this subsection uses homogenous covariance in every dimension.  For the case when $d= 10$ and $\phi = .75$, \figref{fig:smooth_compare} compares the average root mean squared prediction error (RMSPE) resulting from the design strategies computed through $1000$ Monte Carlo samples on $[0,1]^{10}$.     Note if $N>3000$, the RMSPEs for the space-filling designs were not recorded due to numerical instability when inverting the large covariance matrix.

\Figref{fig:smooth_compare}
indicates sparse grid designs yield superior performance to lattice designs.  The results also demonstrate the similarity of the sparse grid designs and space-filling designs in cases of the existence of at least one derivative.  However, sparse grid designs appear inferior to the space-filling designs if the sample path has almost surely no differentiability.
\subsection{Comparison via deterministic functions} \label{sec:deter_fun}
This section will compare the performance of sparse grid designs and space-filling designs on a set of deterministic test functions.  For both methods, we assume the mean and covariance structures of the deterministic functions are unknown and use the \emph{MLE-predictor}.  For $\mu$, we use a constant mean structure, $\mu(\bm x) = \beta$, and for the covariance function we use a scaled Mat\'{e}rn with $\nu = 5/2$ and single lengthscale parameter $\phi$ for all dimensions $i$.     This analysis will report the median absolute prediction error, which is more robust to extreme observations compared to the mean square prediction error.  The median absolute prediction error will be estimated by the sample median of the absolute prediction error at $1000$ randomly selected points in the input space. We consider the following functions: Franke's function \citep{franke1982scattered}, the Borehole function \citep{morris1993bayesian}, the product peak function given by  $y(\bm x) = \prod_{i=1}^{d} (1+10(x^{(i)}-1/4)^2)^{-1}$, the corner peak function given by \[y(\bm x) = \left(1+d^{-1}\sum\nolimits_{i=1}^{d} x^{(i)}\right)^{-d-1},\] and the Rosenbrock function given by
\[y(\bm x) = 4 \sum\nolimits_{i=1}^{d-1}  (x^{(i)}-1)^2 + 400 \sum\nolimits_{i=1}^{d-1} ((x^{(i+1)}-.5) -2(x^{(i)}-.5)^2)^2.\]
With the exception of the Borehole function, all domains are $X= [0,1]^d$ (the Borehole function was scaled to the unit cube).  For the space-filling designs, designs sizes were restricted to cases where memory constraints in MATLAB were not violated on the author's computer.

\begin{figure}[tbp]
\begin{center}
\includegraphics[width=6.5 in]{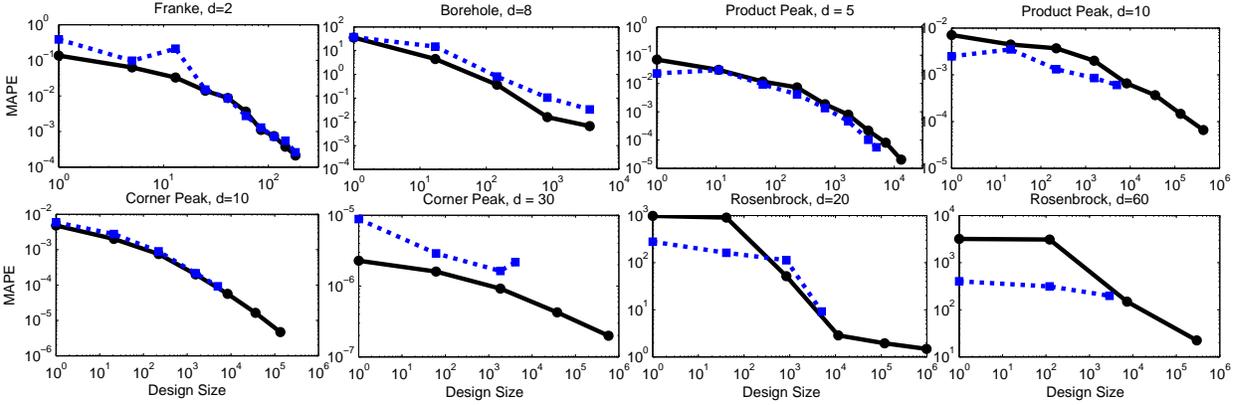}
\caption{Median absolute prediction errors (MAPE) of the MLE-predictor from \secref{sec:gen_prediction} with sparse grid designs (circles, solid line) and space-filling designs (squares, dashed line).}
\label{fig:compare_deterministic}
\end{center}
\end{figure}

\Figref{fig:compare_deterministic}
presents the results of the study. Most functions were similarly estimated using either design strategy.  While Franke's function has significantly more bumps and ridges compared to the other functions, making it more difficult to estimate, good prediction of Franke's function based on few observations is possible because the input to the function is located in a $2$ dimensional space.  At the other extreme, while the corner peak function is smooth, estimating the function when $d=30$ is a very challenging task.  Using a space-filling design of size $4000$ does not do an adequate job of estimating the function as it produces median absolute prediction error of about $10$ times more than the best that can be achieved using a sparse grid design with a much larger design size.  Similar effects are seen when attempting to estimate the Rosenbrock function in $60$ dimensions.

\begin{figure}[tbp]
\begin{center}
\includegraphics[width=3 in]{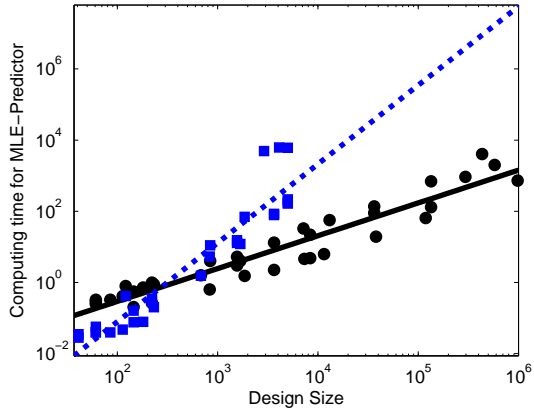}
\caption{Computation time (in seconds) needed to find the MLE-predictor from \secref{sec:gen_prediction} using the proposed method for sparse grid experimental designs (circles) and the traditional method with space-filling designs (squares). The solid line (sparse grid designs) and the dashed line (space-filling designs) represent least squares fits of the model $\log \text{computational time} = \beta_0 + \beta_1 \log N$ to the respective data.  }
\label{fig:comp_time}
\end{center}
\end{figure}

\Figref{fig:comp_time}
compares the computational time needed to find both $\hat{\theta}$ and the weights $\bm w$ using both the traditional method and the proposed method for the MLE-predictors used to produce \figref{fig:compare_deterministic}.  The method to find the MLE-predictor was described in \secref{sec:gen_prediction}.  There was three cases where the cost of the traditional algorithm with a design size of less than $5000$ was more than the proposed algorithm with a design size of nearly a million.  While the design sizes attempted for the traditional algorithm  were limited for memory and numerical stability reasons, some extrapolation emphasizes the problem with using the traditional algorithm on experiments with huge sample sizes.   A sample size of a million points would require roughly $10^7$ seconds, or $115.7$ days,  to find the MLE-predictor.  By using a sparse grid design, we are able to compute the MLE-predictor based on a million observations in a fraction of that time, about 15 minutes ($770$ seconds).
\section{Discussion} \label{sec:conc}
The proposed sparse grid designs deviate from the traditional space-filling framework and utilize lattice structures to construct efficient designs.  Sparse grid designs appear to be competitive with common space-filling designs in terms of prediction, but space-filling designs appear to outperform sparse grid designs in simulations where the underlying function has no differentiability.  Based on the discussions at the end of \secref{sec:mse_comp}, these early results may extend to cases that can be classified as rough functions.

Sparse grid designs are an enormously flexible framework and this work has not yet realized their full potential.
A topic not discussed at length in this work are \emph{optimal} sparse grid designs, which might be able to close any small performance gaps between sparse grid and space-filling designs.  Optimality is dictated by the choice of design criteria, which has previously focused on distance measures such as the minimum distance between any two design points.  When the sample size grows large, this can become an expensive metric to compute as it requires $\mathcal{O} (N^2)$ arithmetic operations.  Therefore, using the shortcut calculations for integrated prediction error, see equation (\ref{eq:sg_error}), or maximum entropy, see \theref{thm:det}, might be faster criteria to compute (see \cite{sacks1989design} for more information on these criteria).

A problem not yet solved using the proposed designs occurs when the covariance function is not separable.
The study of these situations merits more work. As an example, if the sample path contains distinct areas with differing behavior, the assumption of local separability might be a more apt modeling strategy.  Using only local separability assumptions, methods similar to \cite{gramacy2008bayesian} could be employed with local sparse grid designs that study heterogeneous sections of the function.
\section*{Acknowledgments}
Special thanks go to C.F.J. Wu, V.R. Joseph, R. Tuo, J. Lee, an anonymous associate editor, and anonymous reviewers for their inspirations and insights on this current revision.  This work was partially supported by NSF grant CMMI-1030125 and the ORACLE summer fellowship from Oak Ridge National Laboratories.

\appendix
\section*{Appendices}
One notational difference between the body of the paper and these appendices:  Since the proofs for theorems 1 and 2 are demonstrated through induction by treating the level of construction $\eta$ and the dimension $d$ as variables, we use the indexing $(\eta,d)$ for the design $\mathcal{X}_{SG}(\eta,d)$,  the design size $N_{SG}(\eta,d)$, the index sets $\mathbb{J}(\eta,d)$ and $\mathbb{P}(\eta,d)$, and the covariance matrix $\bm \Sigma(\eta,d)$.  Also, the symbol $\setminus$ means `set-minus', i.e. $A \setminus B$ is the elements in $A$ that are not in $B$.
\section{Component designs used for the sparse grid designs in this work} \label{app:componet}
The sparse grid design in \figref{fig:grid_compare}, subplot c, was created where $d=2$, $\eta = 6$ and  $\mathcal{X}_{i,1}, \mathcal{X}_{i,2}\setminus \mathcal{X}_{i,1},\mathcal{X}_{i,3}\setminus \mathcal{X}_{i,2}$, $\mathcal{X}_{i,4}\setminus \mathcal{X}_{i,3}$  and $\mathcal{X}_{i,5}\setminus \mathcal{X}_{i,4}$ are $\{.5\},\{0,1\},\{.25,.75\},\{.375,.625\}$ and $\{.125,.875\}$ respectively for $i=1$ and $2$.

The sparse grid design in \figref{fig:grid_exmp} was created with component designs such that $\mathcal{X}_{i,1}$, $ \mathcal{X}_{i,2}\setminus \mathcal{X}_{i,1}$, $\mathcal{X}_{i,3}\setminus \mathcal{X}_{i,2}$, $\mathcal{X}_{i,4}\setminus \mathcal{X}_{i,3}$, $\mathcal{X}_{i,5}\setminus \mathcal{X}_{i,4}$, $\mathcal{X}_{i,6} \setminus \mathcal{X}_{i,5}$, and $\mathcal{X}_{i,7} \setminus \mathcal{X}_{i,6}$ are $\{.5\}$, $\{.125,.875\}$, $\{.25,.75\}$, $\{0,1\}$, $\{.375,.625\}$, $\{0.1875, 0.8125\},$ and $\{0.0625,0.9375\}$ respectively for all $i$.  These component designs were chosen through an ad-hoc method, but are essentially based on maintaining good spread of points as $\eta$ increases.

The component designs used in \figref{fig:grid_exmp} are used to construct higher dimensional designs used in \secref{sec:simu_comp}.
\section{Proof that \algref{alg:w_comp} produces correct $\bm w$}\label{app:BLUP_Lin}
The correctness of $\bm w$ produced by \algref{alg:w_comp} is difficult to understand without the use of linear operators, therefore we will rephrase $\hat{y} (\bm x_0)$ discussed in \secref{sec:GPmodelandprediction} in terms of a linear operator.  Let $\mathcal{F}$ be a function space of functions that map $X$ to  $\mathbb{R}$.  Let $\mathcal{P}: \mathcal{F} \rightarrow \mathbb{R}$ be a \emph{predictor operator} with respect to $\mathcal{X}=\{\bm x_1, \ldots, \bm x_N\}$ if $\mathcal{P} f =\sum_{k=1}^N q_k f (\bm x_k)$ where $q_k \in \mathbb{R}$.
The following definition explains an \emph{optimal} predictor operator.
\begin{defin}
A predictor operator $\mathcal{P}$ is termed \emph{optimal} with respect to $\bm x_0$ and $\mathcal{X}=\{\bm x_1, \ldots, \bm x_N\}$   if $\mathcal{P} f =\sum_{k=1}^N q_k f (\bm x_k) $ and
\[\{q_1,\ldots,q_N\} = \operatorname{argmin}_{\{\alpha_1,\ldots,\alpha_N\} \in \mathbb{R}^N} \mathbb{E} \left(\mu(\bm x_0) +\sum_{k=1}^N \alpha_k [Y(\bm x_k )-\mu(\bm x_k)]- Y(\bm x_0)\right)^2.\]
\end{defin}
A predictor operator $\mathcal{P}$ is termed optimal because
\begin{equation}
\hat{y} (\bm x_0) = \mu(\bm x_0)+ \mathcal{P}[ y-\mu], \nonumber
\end{equation}
is the \emph{best linear unbiased predictor} of $y(\bm x_0)$ given the observations $\bm Y = \bm y$ when $\mathcal{P}$ is optimal with respect to $\bm x_0$ and $\mathcal{X}$ \citep{santner2003design}.

In general, the optimal predictor operator is when $q_k$ is the $k$th element in $\bm \sigma^\mathsf{T}(\bm x_0) \bm \Sigma^{-1}$.  There are cases where the predictor operator is unique.  Therefore, we only need to show a clever form of the optimal predictor operator that agrees with the  $\bm w$ produced by \algref{alg:w_comp} to complete our argument.

Now we define a sequence, $j = 0,1,2,\ldots$, of predictor operators, $\mathcal{P}_{i,j}$, for each dimension $i$.  These are the optimal predictor operators with respect to $x_0^{(i)}$ and $\mathcal{X}_{i,j}$ when the dimension of the input is $1$ and the covariance function is $C_i$.

To find the desired form of the optimal predictor operator with respect to sparse grid designs, one could guess that the quadrature rule of \cite{smolyak1963quadrature} will be of great use.  In our terms, the \citeauthor{smolyak1963quadrature} quadrature rule can be interpreted as the predictor operator
\begin{equation} \label{eq:sg_pred}
\mathcal{P} (\eta,d) = \sum_{{ \vec{j}} \in \mathbb{J}(\eta,d)} \bigotimes_{i = 1}^d \mathcal{P}_{i,j_i} - \mathcal{P}_{i,j_i-1},
\end{equation}
where the $\otimes$ symbol for linear operators is the tensor product.  .

While this form of $\mathcal{P} (\eta,d)$ is known, the optimality of $\mathcal{P} (\eta,d)$ in the situation discussed has not yet to been proved to the author's knowledge.  \cite{wasilkowski1995explicit} study the case where $\mathcal{X}_{i,j} = \mathcal{X}_{k,j}$ for all $i$ and $k$.  They show an optimality property with respect to an $L_\infty$ norm, which they term \emph{worst case}.  \citeauthor{wasilkowski1995explicit} go on to state in passing that one could verify that (\ref{eq:sg_pred}) is mean of the predictive distribution and therefore optimal in our setting, but they do not demonstrate it in that work.  Here, we formally state and demonstrate this result.
\begin{thm} \label{thm:optimal_sg}
The predictor operator $\mathcal{P} (\eta,d)$ is optimal with respect to $\bm x_0$ and $\mathcal{X}_{SG}(\eta,d)$. Furthermore, $\mathcal{P} (\eta,d)$ can be written in the form
\begin{equation}
\mathcal{P}(\eta,d) = \sum_{{ \vec{j}}\in \mathbb{P}(\eta,d)} a(\vec{j}) \bigotimes_{i = 1}^d \mathcal{P}_{i,j_i}, \label{eq:simplified_pred}
\end{equation}
where $a(\vec{j}) = \left(-1\right)^{\eta - {\scriptscriptstyle|\vec{j}|}} {d-1 \choose \eta-{\scriptscriptstyle|\vec{j}|}}$ and $\mathbb{P}(\eta,d)  = \left\{ \vec{j} \in \mathbb{N}^d | \max(d,\eta-d+1) \leq \sum_{i=1}^d j_i \leq \eta \right\}$.
\end{thm}

The different statements of (\ref{eq:sg_pred}) and (\ref{eq:simplified_pred}) are important to note.  The predictor operator in (\ref{eq:sg_pred}) is theoretically intuitive as it geometrically explains how we maintain orthogonality as $\eta$ grows and allows for the subsequent proof.   However, if we were to attempt to use (\ref{eq:sg_pred}) directly, each term in the sum would require us to sum $2^d$ terms  after expansion, which may temper any computational advantages the lattice structure yields.  \citeauthor{wasilkowski1995explicit} show that (\ref{eq:sg_pred}) can be written of the form (\ref{eq:simplified_pred}).  This result is simply an algebraic manipulation and requires no conditions regarding optimality, but the result allows us to easily use (\ref{eq:sg_pred}).

The fact that (\ref{eq:simplified_pred}) is the optimal predictor operator verifies that \algref{alg:w_comp} produces correct $\bm w$.

\subsection*{Proof of \theref{thm:optimal_sg}}
\begin{proof}
Let
\[\mathcal{P}(\eta,d) f =\sum_{k=1}^N q_k f (\bm x_k)\] where $q_k \in \mathbb{R}$ and $\{\bm x_1, \ldots, \bm x_N\} = \mathcal{X}_{SG}(\eta,d)$.  We need to show that
\[\{q_1,\ldots,q_N\} = \operatorname{argmin}_{\{\alpha_1,\ldots,\alpha_N\} \in \mathbb{R}^N} \mathbb{E} \left(\mu(\bm x_0) +\sum_{k=1}^N \alpha_k[ Y(\bm x_k )-\mu(\bm x_k)]  - Y(\bm x_0)\right)^2.\]
Since $\mathbb{E} (Y(\bm x) - \mu (\bm x) ) =0$, the objective function is minimized when the covariance between $\sum_{k=1}^N \alpha_k Y(\bm x_k )  - Y(\bm x_0)$ and values of $Y$ at all points in $\mathcal{X}_{SG}(\eta,d)$ is $0$.  Thus, we need to show that
\begin{equation}
\operatorname{cov}(\mathcal{P} (\eta,d) Y - Y(\bm x_0), Y(\bm x_k)) =0,\nonumber
\end{equation}
for all $\bm x_k \in \mathcal{X}_{SG}(\eta,d)$.

If $d=1$, the theorem is clearly true for all $\eta \geq d$.  Assume that the theorem is true for $d-1$ and all $\eta\geq d-1$; we will show that it is true for $d$ and $\eta$.  This demonstrates the result by an induction argument.

We have that
\begin{equation}
\operatorname{cov}(\mathcal{P} (\eta,d) Y - Y(\bm x_0 ), Y(\bm x_k))  = - C(\bm x_0, \bm x_k)+\mathcal{P}  (\eta,d) \mathbb{E} \left[\{Y(\bm x_k) -\mu(\bm x_k)\} (Y-\mu)\right].\label{eq:app_cov}
\end{equation}
Observe that
\begin{align}
\mathcal{P}  (\eta,d) \mathbb{E} \left[\{Y(\bm x_k) -\mu(\bm x_k)\} (Y-\mu)\right]&= \mathcal{P} (\eta,d)  C\left(\cdot, \bm x_k\right) \nonumber\\
&= \sum_{{ \vec{j}} \in \mathbb{J}(\eta,d)} \bigotimes_{i=1}^d \mathcal{P}_{i,j_i}C_i \left(\cdot, x_k^{(i)}\right)  - \mathcal{P}_{i,j_i-1} C_i \left(\cdot, x_k^{(i)}\right)  \nonumber\\
&= \sum_{{ \vec{j}} \in \mathbb{J}(\eta-1,d-1)} \prod_{i=1}^{d-1} \mathcal{P}_{i,j_i}C_i \left(\cdot, x_k^{(i)}\right)  - \mathcal{P}_{i,j_i-1} C_i \left(\cdot, x_k^{(i)}\right)  \nonumber\\
& \cdot \sum_{j_d =1}^{\eta-|{ \vec{j}}|} \mathcal{P}_{d,j_d} C_d \left(\cdot, x_k^{(d)}\right)  - \mathcal{P}_{d,j_d-1} C_d  \left(\cdot, x_k^{(d)}\right). \label{eq:expand_one}
\end{align}

Since $\mathcal{P}_{i,j}$ is the optimal predictor operator with respect to $x_0^{(i)}$ and $\mathcal{X}_{i,j}$, $\mathcal{P}_{i,j} C_i \left(\cdot, x\right) - C_i(x_0^{(i)},x) = 0$ if $x \in \mathcal{X}_{i,j}$. Let
\[\mathbb{K} = \{ \vec{j}| \bm x_k \in \mathcal{X}_{i,j_1} \times \mathcal{X}_{2,j_2}  \times \cdots \times \mathcal{X}_{d,j_d},   \vec{j} \in \mathbb{J}(\eta,d)\},\]
and let $\vec{a} = \operatorname{argmin}_{{ \vec{j}} \in \mathbb{K}} |\vec{j}|$.
Since sparse grid designs have nested component designs, if  $j_i \geq a_i$, then $\mathcal{P}_{i,j_i}C_i \left(\cdot, x_k^{(i)}\right) = \mathcal{P}_{i,j_1+1} C_i \left(\cdot, x_k^{(i)}\right) = C_i(x_0^{(i)},x_k^{(i)}),$
since $x_k^{(i)} \in \mathcal{X}_{i,j_i} \subset \mathcal{X}_{i,j_i+1}$.   This implies if $\vec{j} \not \leq \vec{a}$, then $\prod_{i=1}^{d} \left(\mathcal{P}_{i,j_i}C_i \left(\cdot, x_k^{(i)}\right)  - \mathcal{P}_{i,j_i-1} C_i \left(\cdot, x_k^{(i)}\right) \right) = 0.$ Then (\ref{eq:expand_one}) can be rewritten as
\begin{multline}
\mathcal{P}  (\eta,d) \mathbb{E} \left[\{Y(\bm x_k) -\mu(\bm x_k)\} (Y-\mu)\right]  =\\
\sum_{{ \vec{j}} \in \mathbb{J}(\eta-1, d-1)} \prod_{i=1}^{d-1} \left(\mathcal{P}_{i,j_i}C_i \left(\cdot, x_k^{(i)}\right)  - \mathcal{P}_{i,j_i-1} C_i \left(\cdot, x_k^{(i)}\right) \right)  \\
\cdot \sum_{j_d =1}^{\max(\eta-a_1 -\cdots  - a_{d-1}, \eta -|{ \vec{j}}|)} \mathcal{P}_{d,j_d}  C_d \left(\cdot, x_k^{(d)}\right)  - \mathcal{P}_{d,j_d-1}  C_d \left(\cdot, x_k^{(d)}\right)\label{eq:expand_two}
\end{multline}
Also, if $j_d>a_d$ then $\mathcal{P}_{d,j_d} C_d \left(\cdot, x_k^{(d)}\right)  - \mathcal{P}_{d,j_d-1} C_d \left(\cdot, x_k^{(d)}\right) = 0$  and \[\sum_{i=1}^d a_i \leq \eta \Rightarrow a_d \leq max(\eta  - {a_1 -\cdots  - a_{d-1}}, \eta  - {j_1 -\cdots  - j_{d-1}}),\]
which implies
\begin{align}
\sum_{j_d =1}^{\max(\eta-a_1 -\cdots  - a_{d-1}, \eta -|{ \vec{j}}|)} \mathcal{P}_{d,j_d} C_d \left(\cdot, x_k^{(d)}\right)  - &\mathcal{P}_{d,j_d-1} C_d \left(\cdot, x_k^{(d)}\right)    \nonumber\\
&=\sum_{j_d =1}^{a_d} \mathcal{P}_{d,j_d} C_d \left(\cdot, x_k^{(d)}\right)  - \mathcal{P}_{d,j_d-1} C_d \left(\cdot, x_k^{(d)}\right) \nonumber\\
& = C_d \left(x_0^{(d)}, x_k^{(d)}\right).\label{eq:d_expansion}
\end{align}

Plugging (\ref{eq:d_expansion}) into (\ref{eq:expand_two}) yields
\begin{multline}
\mathcal{P}  (\eta,d) \mathbb{E} \left[\{Y(\bm x_k) -\mu(\bm x_k)\} (Y-\mu)\right]=\\ C_d \left(x_0^{(d)}, x_k^{(d)}\right) \cdot \sum_{{ \vec{j}} \in \mathbb{J}(\eta-1,d-1)} \prod_{i=1}^{d-1} \mathcal{P}_{i,j_i}C_i \left(\cdot, x_k^{(i)}\right)  - \mathcal{P}_{i,j_i-1} C_i \left(\cdot, x_k^{(i)}\right). \nonumber
\end{multline}
By the induction assumption, the theorem is true for $d-1$, which means that for $\eta-1$ and $d-1$, (\ref{eq:app_cov}) is equal to zero. Therefore,
\[\sum_{{ \vec{j}} \in \mathbb{J}(\eta-1,d-1)} \prod_{i=1}^{d-1} \mathcal{P}_{i,j_i}C_i \left(\cdot, x_k^{(i)}\right)  - \mathcal{P}_{i,j_i-1} C_i \left(\cdot, x_k^{(i)}\right) = \prod_{i=1}^{d-1} C_i \left(x_0^{(i)}, x_k^{(i)}\right). \]
This gives us the desired result for $d$,
\begin{align}
\mathcal{P}  (\eta,d) \mathbb{E} \left[\{Y(\bm x_k) -\mu(\bm x_k)\} (Y-\mu)\right]  = \prod_{i=1}^d  C_i \left(x_0^{(i)}, x_k^{(i)}\right) =C(\bm x_0, \bm x_k). \nonumber
\end{align}
Inserting this into (\ref{eq:app_cov}) yields the major result that (\ref{eq:sg_pred}) is the optimal predictor operator.

In \cite{wasilkowski1995explicit}, they demonstrate through combinatorial relations and algebraic manipulations that (\ref{eq:sg_pred}) can be simplified to (\ref{eq:simplified_pred}).
\end{proof}
\section{Proof that (\ref{eq:sg_error}) is the MSPE} \label{app:SG_MSPE}
Due to \theref{thm:optimal_sg},
\begin{align}
\mathbb{E} \left(\hat{Y} (\bm x_0)-Y(\bm x_0)\right)^2 &= \operatorname{var}(Y (\bm x_0)) - \mathcal{P} (\eta,d) \mathbb{E} \left[\{Y (\bm x_0) - \mu (\bm x_0)\} (Y-\mu)\right] \nonumber\\
& = C(\bm x_0, \bm x_0)
-\sum_{{ \vec{j}} \in \mathbb{J}(\eta,d)} \prod_{i=1}^d \mathcal{P}_{i,j}C_i(\cdot,x_0^{(i)}) -\mathcal{P}_{i,j-1}C_i(\cdot,x_0^{(i)}). \nonumber
\end{align}
Because $\mathcal{P}_{i,j}$ is the optimal predictor operator in one dimension with respect to $x_0^{(i)}$ and $\mathcal{X}(\eta,d)$, we have
\begin{equation}
\mathbb{E} \left(\hat{Y} (\bm x_0)-Y(\bm x_0)\right)^2 = \operatorname{var}(Y (\bm x_0)) - \sum_{{ \vec{j}} \in \mathbb{J}(\eta,d)} \prod_{i=1}^d \Delta_{i,j_i}, \nonumber
\end{equation}
where $\Delta_{i,j}$ is defined in \secref{sec:sg_infr}.

Lastly, we have that since $\hat{Y}(\cdot)$ is an affine map from $\bm Y$ and $Y(\cdot)$ follows a Gaussian process, $\hat{Y}(\bm x_0) - Y(x_0)$ and $\bm Y$ are jointly multivariate normal.  By \theref{thm:optimal_sg}, there is $0$ covariance between them.  Therefore $\hat{Y}(\bm x_0) - Y(x_0)$ is independent of $\bm Y$ and we can condition the expectation on the left-hand-side on $\bm Y = \bm y$ without affecting the right-hand-side.
\section{Proof of \theref{thm:det}} \label{app:det_proof}
\begin{proof}
If $d = 1$, the theorem is clearly true for all $\eta\geq d$.  We now prove this result by induction.  Assume the theorem is true for $d-1$ and all $\eta\geq d-1$.

To demonstrate this result, we require the use of the Schur complement.  Let $\bm M = \left[\bm A, \bm B; \bm B^\mathsf{T}, \bm C\right]$. The Schur complement of $\bm M$ with respect to $\bm A$, expressed $\bm M \left/ \bm A \right.$, is defined by $\bm C - \bm B^\mathsf{T} \bm A^{-1} \bm B$ (if $\bm A$ is invertible).  The determinant quotient property of the Schur complement is $|\bm M / \bm A| = |\bm M ||\bm A |^{-1}$.   The theorem can be rewritten as
\begin{equation}
|\bm \Sigma(\eta,d)| = \prod_{{ \vec{j}} \in \mathbb{J}(\eta,d)} \prod_{i=1}^d \left|\bm S_{i,j_i}\left/\bm  S_{i, j_i -1}\right|\right.^{\prod_{k \neq i} \#\mathcal{X}_{k,j_k}-\#\mathcal{X}_{k,j_k-1}} . \nonumber
\end{equation}
We also require following result:
\begin{equation}
|\bm A \otimes \bm B| = |\bm A|^m |\bm B|^n, \label{eq:det_rule}
\end{equation}
where $\bm A$ and $\bm B$ are $n \times n$  and $m \times m$ sized matrices, respectively.

We will use the notation $\bm \Sigma(\eta,d; \mathcal{X}_0)$ to denote the submatrix of $\bm \Sigma(\eta,d)$ with respect to the elements which correspond to $\mathcal{X}_0 \subset \mathcal{X}$.  Expanding the term $\left|\bm \Sigma(\eta,d)\right|$ with respect the quotient property
\begin{align}
\left|\bm \Sigma(\eta,d)\right|  =& \left|\bm \Sigma(\eta,d) \left/ \bm \Sigma(\eta,d;\mathcal{X}_{SG} (\eta-1 ,d-1)\times \mathcal{X}_{d,1}\setminus \mathcal{X}_{d,0})\right.\right|\nonumber\\
& \left|\bm \Sigma(\eta,d;\mathcal{X}_{SG} (\eta-1 ,d-1)\times \mathcal{X}_{d,1}\setminus \mathcal{X}_{d,0})\right|.\label{eq:det_exp}
\end{align}
Let \[\bm Q = \bm \Sigma(\eta,d) \left/ \bm \Sigma(\eta,d;\mathcal{X}_{SG} (\eta-1 ,d-1)\times \mathcal{X}_{d,1}\setminus \mathcal{X}_{d,0})\right. .\] Now, observe the elements of $\bm Q$ that correspond to $\mathcal{X}_{SG} (\eta-2 ,d-1) \times \mathcal{X}_{d,2}\setminus \mathcal{X}_{d,1}$,
\begin{align}
\bm Q (\mathcal{X}_{SG} (\eta-2 ,d-1) \times \mathcal{X}_{d,2}\setminus \mathcal{X}_{d,1}) &= \bm A - \bm B^\mathsf{T} \bm C^{-1} \bm B, \nonumber
\end{align}
where $\bm A$ is a covariance matrix corresponding to $\mathcal{X}_{SG} (\eta-2 ,d-1)\times \mathcal{X}_{d,2}\setminus \mathcal{X}_{d,1}$, $\bm C$ is a covariance matrix corresponding to $\mathcal{X}_{SG} (\eta-1 ,d-1)\times \mathcal{X}_{d,1}\setminus \mathcal{X}_{d,0}$, and $\bm B$ is the cross covariance.  Since $\mathcal{X}_{SG} (\eta-2 ,d-1) \subset \mathcal{X}_{SG} (\eta-1 ,d-1)$ and $\mathcal{X}_{d,1}\subset \mathcal{X}_{d,2}$,
\begin{align}
\bm Q(\mathcal{X}_{SG} (\eta-2 ,d-1) \times \mathcal{X}_{d,2}\setminus \mathcal{X}_{d,1}) &=    \bm \Sigma\left(\eta-2,d-1\right) \bigotimes \left(\bm S_{d,2} \left/ \bm S_{d,1}\right.\right). \nonumber
\end{align}
So,
\begin{align}
|\bm Q (\mathcal{X}_{SG} (\eta-2 ,d-1) \times \mathcal{X}_{d,2}\setminus \mathcal{X}_{d,1})| &= \left|\bm \Sigma(\eta,d; \mathcal{X}_{SG}(\eta-2 ,d-1)\times \mathcal{X}_{d,2}\setminus \mathcal{X}_{d,1})\right|, \nonumber
\end{align}
which can be used with (\ref{eq:det_exp}) to show
\begin{align}
\bm \Sigma(\eta,d) =& \left|\bm \Sigma(\eta,d) \left/ \bm \Sigma(\eta,d;\mathcal{X}_{SG} (\eta-1 ,d-1)\times \mathcal{X}_{d,1}\setminus \mathcal{X}_{d,0})\right.\left/ \bm \Sigma(\eta,d;\mathcal{X}_{SG} (\eta-2 ,d-1)\times\mathcal{X}_{d,2}\setminus \mathcal{X}_{d,1})\right.\right|\nonumber\\
& \left|\bm \Sigma(\eta,d;\mathcal{X}_{SG} (\eta-1 ,d-1)\times \mathcal{X}_{d,1}\setminus \mathcal{X}_{d,0})\right| \left|\bm \Sigma(\eta,d;\mathcal{X}_{SG} (\eta-2 ,d-1)\times \mathcal{X}_{d,2}\setminus \mathcal{X}_{d,1})\right|.\nonumber
\end{align}
Iterating the expansion to $\eta-d+1$ yields,
\begin{align}
\left|\bm \Sigma(\eta,d)\right|  =& \left|\bm \Sigma(\eta,d) \left/ \bm \Sigma(\eta,d;\mathcal{X}_{SG} (\eta-1 ,d-1)\times \mathcal{X}_{d,1}\setminus \mathcal{X}_{d,0})\right.\right. \nonumber \\
&\left.\left. \left/ \bm \Sigma(\eta,d;\mathcal{X}_{SG} (\eta-2 ,d-1)\times \mathcal{X}_{d,2}\setminus \mathcal{X}_{d,1}) \right.\right.\right. \left/ \cdots \phantom{\left.\left. \left/ \bm \Sigma(\eta,d;\mathcal{X}_{SG} (\eta-1 ,d-1)\times \mathcal{X}_{d,2}\setminus \mathcal{X}_{d,1}) \right.\right.\right.} \right.  \nonumber\\
&\left.\left.\left.\left.  \left/ \bm \Sigma(\eta,d;\mathcal{X}_{SG} (d-1 ,d-1)\times \mathcal{X}_{d,\eta-d+1}\setminus \mathcal{X}_{d,\eta-d})  \right.\right.\right.\right.\right| \nonumber\\
&\prod_{j_d=1}^{\eta-d+1 } \left|\bm \Sigma(\eta,d; \mathcal{X}_{SG} (\eta-j_d ,d-1)\times \mathcal{X}_{d,j_d}\setminus \mathcal{X}_{d,j_d-1} )\right| \nonumber.
\end{align}
The term outside of the product is the Schur complement of a positive definite matrix with itself, which is an empty matrix.  By the Leibniz formula, the determinant is $1$.  Therefore,
\begin{align}
\left|\bm \Sigma(\eta,d)\right|  =\prod_{j_d=1}^{\eta-d+1 } \left|\bm \Sigma(\eta,d; \mathcal{X}_{SG} (\eta-j_d ,d-1)\times \mathcal{X}_{d,j_d}\setminus \mathcal{X}_{d,j_d-1} )\right| \nonumber.
\end{align}

With (\ref{eq:det_rule}), we have
\[\left|\bm \Sigma(\eta,d)\right|  = \prod_{j_d=1}^{\eta-d+1 } \left|\bm S_{d,j_d} \left/\bm  S_{d,j_d -1}  \right.\right|^{N_{SG}(\eta-j_d,d-1)} |\bm \Sigma(\eta-j_d,d-1)|^{\#\mathcal{X}_{d,j_d}-\#\mathcal{X}_{d,j_d-1}}.\]
And by (\ref{eq:num_points}) and the induction assumption
\begin{align}
|\bm \Sigma(\eta,d)| &= \prod_{j_d=1}^{\eta-d+1 } \prod_{{ \vec{j}} \in \mathbb{J}(\eta-j_d, d-1)}  \left|\bm S_{d,j_d} \left/ \bm  S_{d,j_d -1}\right.\right|^{\prod_{k \neq d} \#\mathcal{X}_{k,j_k}-\#\mathcal{X}_{k,j_k-1}} \prod_{i=1}^{d-1} \left|\bm S_{i,j_i} \left/\bm  S_{i,j_i -1}\right.\right|^{\prod_{k \neq i} \#\mathcal{X}_{k,j_k}-\#\mathcal{X}_{k,j_k-1}},
 \nonumber\\
&= \prod_{{ \vec{j}} \in \mathbb{J}(\eta, d)}  \prod_{i=1}^{d} \left|\bm S_{i,j_i} \left/\bm  S_{i,j_i -1}\right.\right|^{\prod_{k \neq i} \#\mathcal{X}_{k,j_k}-\#\mathcal{X}_{k,j_k-1}}, \nonumber
 \end{align}
which demonstrates the result.
\end{proof}
\bibliographystyle{./bib/asa}
\bibliography{./bib/SG_ref}
\end{document}